\documentclass{jfm}
\usepackage{graphicx}
\usepackage{natbib}

\ifCUPmtlplainloaded \else
  \checkfont{eurm10}
  \iffontfound
    \IfFileExists{upmath.sty}
      {\typeout{^^JFound AMS Euler Roman fonts on the system,
                   using the 'upmath' package.^^J}%
       \usepackage{upmath}}
      {\typeout{^^JFound AMS Euler Roman fonts on the system, but you
                   dont seem to have the}%
       \typeout{'upmath' package installed. JFM.cls can take advantage
                 of these fonts,^^Jif you use 'upmath' package.^^J}%
      }
  \else
  \fi
\fi

\ifCUPmtlplainloaded \else
  \checkfont{msam10}
  \iffontfound
    \IfFileExists{amssymb.sty}
      {\typeout{^^JFound AMS Symbol fonts on the system, using the
                'amssymb' package.^^J}%
       \usepackage{amssymb}%
         \let\leq=\leqslant
         
      }{}
  \fi
\fi

\ifCUPmtlplainloaded \else
  \IfFileExists{amsbsy.sty}
    {\typeout{^^JFound the 'amsbsy' package on the system, using it.^^J}%
     \usepackage{amsbsy}}
    {}
\fi

\newsavebox{\astrutbox}
\sbox{\astrutbox}{\rule[-5pt]{0pt}{20pt}}

\title[Exact spiral solution of 3D Euler equations]{An exact axisymmetric spiral solution of the incompressible 3D Euler equations}

\author[L. Sun]%
{L\ls I\ls A\ls N\ls G\ns  S\ls U\ls N$^1$$^2$%
  \thanks{Corresponding Author: Liang Sun, email: sunl@ustc.edu.cn;\break sunl@ustc.edu.}
}

\affiliation{
$^1$School of Earth and Space Sciences, University of Science and
Technology of China, Hefei, 230026, China.\\[\affilskip]
$^2$Dept. of Modern Mechanics, University of Science and
Technology of China, Hefei, 230026, China.}

\date{\today}

\date{\today, and in revised form ??}

\begin{document}

\maketitle

\begin{abstract}
Spiral structure is one of the most common structures in the
nature flows. A general steady spiral solution of incompressible
inviscid axisymmetric flow was obtained analytically by applying
separation of variables to the 3D Euler equations. The solution,
depending on 3 parameters, describes the spiral path of the fluid
material element on the Bernoulli surface, whereas some new exact
solutions were obtained to be bounded within the whole region. The
first one is a continued typhoon-like vortex solution, where there
are two intrinsic length scales. One is the radius of maximum
circular velocity $r_m$, the other is the radius of the vortex
kernel $r_k=\sqrt{2}r_m$. The second one is a multi-planar
solution, periodically in $z$-coordinate. Within each layer, the
solution is a umbrella vortex similar to the first one. The third
one is also a multi-planar solution in $z$-coordinate. In each
layer, it is a combination of two independent solutions like the
Rankine vortex, which is also finite but discontinued for either
vertical or horizontal velocity. The fourth one is a
multi-paraboloid vortex solution finite for $z$-coordinate but
infinite for $r$-coordinate. Besides, some classical simple
solutions (Rankine vortex, Batchelor vortex, Hill spherical
vortex, etc.) were also obtained. The above explicit solutions can
be applied to study the radial structure of the typhoon, tornados
and mesoscale eddies. Both the solutions and approaches used here
could also be applied to other complex flows by the Navier-Stokes
equations.

\end{abstract}

\section{Introduction}
The exact solutions of the Euler equations for inviscid flow are
quite important for understanding how the real fluid will flow.
But it is a big problem for solving the Euler equations due to the
nonlinearity. Within 2D (two-dimensional) context, the complex
potential can effectively solve the irrotational flow by turning
the Euler equations to linear ones. For the 2D inviscid flow with
vorticity, the vorticity-stream formulation can simplify the
problem to a generalized Helmholtz equation
\cite[]{Batchelor1967}. Consequently, some special explicit
solutions were obtained after this simplification
\cite[]{WangCY1991,Saffman1992,MajdaBook2002,WuJZbook2006}. And
\cite{LouSY2007} also obtained some non-steady solutions from the
steady solutions by the general symmetry group theorem.

Within 3D (three-dimensional) flow context, the problem becomes
more complex, especially for the 3D flow with vorticity and
swirling. The nonlinear advection term make the equations hardly
to be solved. If the nonlinear advection term vanishes everywhere
at certain conditions, the flow is called the Beltrami flow. Thus
the Euler equations could be simplified to linear equations, and
some exact solutions might be obtained. The known limited exact
solutions of the Euler and Navier-Stokes equations were reviewed
in detail by \cite{WangCY1991,Schtern1999}. Such solutions from
simplicity to complexity include concentric flows, parallel flows,
linear flows, the generalized Beltrami flows in axisymmetric case
\cite[]{WangCY1991}. Among them, the exact solutions for
generalized Beltrami flows are the most complex, and they describe
swirling flows of widespread technological importance.

Within 3D axisymmetric context, the simplest Beltrami flow is a
steady swirling flow, where both the radial velocity and azimuthal
velocities are dependent of the radial coordinate $z$, but of the
axial coordinate $r$ \cite[]{Schtern1997}. In this case, the Euler
equations for the radial velocity, the azimuthal velocity and
swirl are decoupled, where the coupling of three velocity
components are arbitrarily constructed rather than derived from
reasonable boundary conditions. The solutions include the Burgers
vortex, the Sullivan vortex
\cite[e.g.][]{WangCY1991,WuJZbook2006}, and a family of general
sink vortices \cite[]{Schtern1997}. Another simple
axisymmetrically Beltrami flow is symmetric sink without swirling.
Some of them are impingement of two rotating streams, rotational
stagnation flow over a plate, the Hill spherical vortex and etc
\cite[]{WangCY1991,WuJZbook2006}. \cite{Moffatt1969} also extended
the Hill spherical vortex to allow for a swirl with nonzero
circumferential velocity that causes a circulation around the
$z$-axis.

Although the above solutions are helpful, most of the symmetric
sink solutions are decoupled and unbounded, except for, e.g., the
Hill spherical vortex and its extension. The decoupled and
unbounded solutions might be of some limited physical interest. To
avoid this arbitrariness and unboundedness,
\cite{Long1958,Long1961} tried first to obtain the coupled
equations for conical similarity swirling vortex, and the Euler
equations can be simplified to the Bragg-Hawthorne equation, or
the Long-Squire equation
\cite[]{Batchelor1967,Saffman1992,Frewer2007}. However, such
equation is generally nonlinear, no exact explicit solution has
been obtained.

In contrast to that, \cite{Frewer2007} used Lie-algebras to
investigate the Bragg-Hawthorne equation. They found out eight
distinct local Bragg-Hawthorne symmetries within five algebras.
However, such symmetries are different and even more symmetries
than the Euler equations do \cite[]{Frewer2007}. To dispel the
inequivalent between them, they need to obtain some additional
equations. Even though, they also find only the simple linear
Bragg-Hawthorne equation can be solved explicitly
\cite[]{Batchelor1967,Frewer2007}.

As an alternative way to avoid the above disadvantage, we
considered to solve the coupled nonlinear Euler equations
following the deviation by \cite{Batchelor1967}, which is the most
complex Beltrami flow with that all the velocity components are
dependent on both the radial coordinate $r$ and the axial
coordinate $z$. Due to the complexity, the general explicit
solution for such flow still lacks even in the recent literatures
\cite[]{MajdaBook2002,WuJZbook2006}. In this study, we solved the
Euler equations by applying separation of variables
\cite[]{SunL2011taml}. As a result, a general exact spiral
solution was presented in~\S\,\ref{sec:modelresults}, some special
explicit solutions were given in~\S\,\ref{sec:sepcialsol}.
Discussion and conclusion were respectively given
in~\S\,\ref{sec:discussion} and in~\S\,\ref{sec:conclusion}.

\section{General solution} \label{sec:modelresults}

We considered the steady solution of the incompressible Euler
equations for axisymmetric flow at present study. It is convenient
to use a cylindrical coordinate system $(r,\theta,z)$ with the
velocity components ($V_r,V_{\theta},V_z$), and all the velocity
components are the functions of $r$ and $z$ but $\theta$, due to
the axisymmetric. As $V_r=V_r(r,z)$, $V_{\theta}=V_{\theta}(r,z)$
and $V_z=V_z(r,z)$, the governing equations, including
mass-conservation and momentum equations, are
\begin{subeqnarray}
\frac{\partial (rV_r)}{\partial r} + \frac{\partial (rV_z)}{\partial z} &=&0\\
V_r\frac{\partial V_{\theta}}{\partial r} + V_z\frac{\partial V_{\theta}}{\partial z}+\frac{ V_r V_{\theta}}{r} &=&0\\
V_r\frac{\partial V_{r}}{\partial r} + V_z\frac{\partial V_{r}}{\partial z}-\frac{ V_{\theta}^2}{r} &=&-\frac{1}{\rho}\frac{\partial p}{\partial r}\\
V_r\frac{\partial V_{z}}{\partial r} + V_z\frac{\partial
V_{z}}{\partial z} &=&-\frac{1}{\rho}\frac{\partial p}{\partial z}
 \label{Eq:Axisflow-ctl}
 %\tag{\ref{Eq:2dmodel_ctl_Horizontal}}
 \end{subeqnarray}

It should be noted that there is no length scale in
Eq.(\ref{Eq:Axisflow-ctl}), thus the solution can be uniformly
stretched by simply multiplying a complex constant.
\cite{Batchelor1967} noted that Eq.(\ref{Eq:Axisflow-ctl}a) and
Eq.(\ref{Eq:Axisflow-ctl}b) can be solved through the introduction
of the axisymmetric stream function $\Psi(r,z)$. We tried to find
the solution of the above equations by separation of variables
$\Psi(r,z)=R(r)H(z)$. One such solution can be written as,
\begin{equation}
V_r = \frac{ R(r)}{ r} H'(z),  \,\, V_{\theta} = \lambda \frac{
R(r)}{ r} H(z), \,\, V_{z} = - \frac{ R'(r)}{r} H(z)
 \label{Eq:Axisflow-velsol-gen}
 %\tag{\ref{Eq:2dmodel_ctl_Horizontal}}
 \end{equation}
where $'$ presents first deviation and $\lambda$ is a complex
constant. In this way, equation (\ref{Eq:Axisflow-ctl}a) and
equation (\ref{Eq:Axisflow-ctl}b) are satisfied automatically. The
path of a fluid material element can be obtained,
\begin{subeqnarray}
\ln r({\theta}) &=& \displaystyle \frac{1}{\lambda}\frac{H'}{H}\theta \\
\Psi(r,z) &=& const.
 \label{Eq:Axisflow-velsol-path}
 %\tag{\ref{Eq:2dmodel_ctl_Horizontal}}
 \end{subeqnarray}
In $r-\theta$ plan, it is a logarithmic spiral ($H'\neq0$), except
for $H'=0$ (a circle). So we called the solution is spiral
solution. In fact, the path is right on a Bernoulli surface given
by Eq.(\ref{Eq:Axisflow-velsol-path}b), on which the
streamfunction is defined \cite[]{Batchelor1967}.

Equation (\ref{Eq:Axisflow-ctl}c) and equation (\ref{Eq:Axisflow-ctl}d) become,
\begin{subeqnarray}
(\frac{R}{r})(\frac{R}{r})'H'^2 - (\frac{R}{r})(\frac{R'}{r})HH''-\frac{\lambda^2 }{r^3}R^2H^2 =-\frac{1}{\rho}\frac{\partial p}{\partial r}  \\
(\frac{R'}{r})^2 HH' - (\frac{R}{r})(\frac{R'}{r})' HH' =
-\frac{1}{\rho}\frac{\partial p}{\partial z}
 \label{Eq:Axisflow-velsol-pressure}
 %\tag{\ref{Eq:2dmodel_ctl_Horizontal}}
 \end{subeqnarray}
Hence, the pressure can be solved from
Eq.(\ref{Eq:Axisflow-velsol-pressure}b),
\begin{equation}
p(r,z)=\frac{1}{2}\rho H^2 [(\frac{R}{r})(\frac{R'}{r})'
-(\frac{R'}{r})^2]-\rho Q(r)
 \label{Eq:Axisflow-presol}
 %\tag{\ref{Eq:2dmodel_ctl_Horizontal}}
 \end{equation}
Substituting Eq.(\ref{Eq:Axisflow-presol}) into
Eq.(\ref{Eq:Axisflow-velsol-pressure}a), it yields
\begin{equation}
\displaystyle (\frac{R}{r})(\frac{R}{r})'H'^2 -
(\frac{R}{r})(\frac{R'}{r})(HH'')-(\frac{\lambda^2 }{r^3}R^2) H^2
= \displaystyle  \frac{1}{2}[(\frac{R'}{r})^2 -
(\frac{R}{r})(\frac{R'}{r})']' H^2+Q'(r)
 \label{Eq:Axisflow-velsol-cross}
 %\tag{\ref{Eq:2dmodel_ctl_Horizontal}}
 \end{equation}
Recall that $R$ and $H$ are independent, we can solve the above
equation. There are three kind of functions for $H(z)$ in
Eq.(\ref{Eq:Axisflow-velsol-cross}). The first trivial one is
$H=1$, any differentiable function $R(r)$ would be the solution,
which we discussed in ~\S\,\ref{sec:sepcialsol-keq0-H=1}. Besides,
it also yields two kind of non-trivial solutions for $H(z)$: a)
both $H'^2$ and $HH''$ are independent to $H^2$, and b) both
$H'^2$ and $HH''$ are proportion to $H^2$. This yields
\begin{subeqnarray}
HH'' &=& \frac{n}{n-1} H'^2, n=1,2,\,\, or, \\
HH'' &=& k^2 H^2, and, \,\, H'^2 =k^2 H^2 \pm 4\alpha\beta k^2.
\,\,
 \label{Eq:Axisflow-velsol-Hequation}
 %\tag{\ref{Eq:2dmodel_ctl_Horizontal}}
 \end{subeqnarray}
where $k$ and $\beta$ are two complex constants. The solutions are
obtained for $H(z)$ and $Q(r)$,
\begin{subeqnarray}
H(z) &=& z^{n} , Q(r)=\displaystyle \frac{1}{2}(\frac{R}{r})^2-(n-1)\int \frac{RR'}{r^3} dr,\,\, or,\\
H(z) &=& \alpha e^{kz}\pm \beta e^{-kz}, Q(r)=\displaystyle
 \pm 2\alpha\beta k^2(\frac{R}{r})^2, \,\,
k^2\neq 0,\,\, \label{Eq:Axisflow-velsol-H}
 %\tag{\ref{Eq:2dmodel_ctl_Horizontal}}
 \end{subeqnarray}
Substitution Eq.(\ref{Eq:Axisflow-velsol-H}) into
Eq.(\ref{Eq:Axisflow-velsol-cross}) to eliminate the function for
$z$, and letting $\gamma=\lambda^2+k^2$, it yields
\begin{equation}
 -\frac{2\gamma}{r^3}R^2 = [(\frac{R'}{r})^2 - (\frac{R}{r})(\frac{R'}{r})']'
 \label{Eq:Axisflow-velsol-Requaion}
 %\tag{\ref{Eq:2dmodel_ctl_Horizontal}}
 \end{equation}
where $k=0$ is for Eq.(\ref{Eq:Axisflow-velsol-H}a) and $k\neq0$
is for Eq.(\ref{Eq:Axisflow-velsol-H}b). The solution of Eq.
(\ref{Eq:Axisflow-velsol-Requaion}) for $\gamma \neq0$ yields,
\begin{equation}
R(r)=ar^2e^{-\frac{1}{8}\gamma r^2}
 \label{Eq:Axisflow-velsol-R,gamma!=0}
 %\tag{\ref{Eq:2dmodel_ctl_Horizontal}}
 \end{equation}
and the solution for $\gamma=0$ yields
\begin{subeqnarray}
R(r) &=& ar^2-b, or\\
R(r) &=& ae^{-cr^2}
 \label{Eq:Axisflow-velsol-R,gamma=0}
 %\tag{\ref{Eq:2dmodel_ctl_Horizontal}}
 \end{subeqnarray}
where $a$, $b$ and $c$ are complex constants. And the solution
Eq.(\ref{Eq:Axisflow-velsol-R,gamma=0}b) is independent to
Eq.(\ref{Eq:Axisflow-velsol-R,gamma=0}a) under the condition of
$\gamma=0$, as the function Eq.(\ref{Eq:Axisflow-velsol-Requaion})
is nonlinear. Accordingly, by integrating
Eq.(\ref{Eq:Axisflow-presol}) with
Eq.(\ref{Eq:Axisflow-velsol-R,gamma!=0}) the pressure is,
\begin{equation}
p(r,z)=\rho H^2 (d-4 R^2)-\rho Q(r)
 \label{Eq:Axisflow-presol-explicit}
 %\tag{\ref{Eq:2dmodel_ctl_Horizontal}}
 \end{equation}

Thus, Eq.(\ref{Eq:Axisflow-velsol-gen}) with
Eq.(\ref{Eq:Axisflow-velsol-H})  and
Eq.(\ref{Eq:Axisflow-velsol-R,gamma!=0}) or
Eq.(\ref{Eq:Axisflow-velsol-R,gamma=0}) give the exact solutions
of the flow velocity. It is obvious that there are three
independent parameters (e.g., $a$, $k$, $\lambda$) in the
solutions. And other new solutions can also be obtained by
combining the different solution at different regions, like that
of the Rankine vortex.

%
%This implies the future works to understanding the interaction
%between waves and background flow from energy transport point of
%view.

\section{Special solutions}  \label{sec:sepcialsol}

As we know, the Euler equations are defined in the real field.
While the above general solutions are defined in the complex
field. If the parameters in the solutions are real, the solutions
are real. If the parameters are complex, we should try to take the
real part of the complex as the solution. However, such approach
can not work in general (some additional conditions are required),
in that the Euler equations are nonlinear. Such kind of property
is common for nonlinear equations, even for a simple nonlinear
algebra equation. This problem is similar to the previously
mentioned one in \cite{Frewer2007}, who also found out not all the
Lie-algebra solutions satisfy the Euler equations. In their
studies, some additional equations required.

For the velocity, the combination of
Eq.(\ref{Eq:Axisflow-velsol-H}) and
Eq.(\ref{Eq:Axisflow-velsol-R,gamma!=0}) or
Eq.(\ref{Eq:Axisflow-velsol-R,gamma=0}) according to
Eq.(\ref{Eq:Axisflow-velsol-gen}) gives the solution. However,
there are too many parameters (including $\lambda$, $\alpha$,
$\beta$, $\gamma$, $k$, $a$, $b$, $c$, $n$) in such solution. Our
further investigations also pointed out that
Eq.(\ref{Eq:Axisflow-ctl}b) holds only for $\lambda$ being a real
number, i.e., $\lambda_i=0$. Hence, we restricted our following
investigations to $\alpha=1$, $\beta=0$, $a=1$ and $b=0$ without
loss of generality. Although such restrictions make the flows to
be much more simple, they are not trivial. In the following
section, we gave some interesting solutions to show how the fluid
flows in such conditions.

\subsection {$k\neq0$}
For $k \neq 0$, $H(z)=e^{kz}$,
$\gamma_r=k_r^2-k_i^2+\lambda_r^2-\lambda_i^2$ and
$\gamma_i=2(k_rk_i+\lambda_r\lambda_i)$, we have the parameters
\begin{subeqnarray}
A_r=(a_rk_r-a_ik_i),&B_r=(a_r\lambda_r-a_i\lambda_i),&C_r=a_r-\frac{1}{8}(a_r\gamma_r-a_i\gamma_i)r^2\\
A_i=(a_rk_i+a_ik_r),&B_i=(a_r\lambda_i+a_i\lambda_r),&C_i=a_i-\frac{1}{8}(a_r\gamma_i+a_i\gamma_r)r^2
 \label{Eq:Axisflow-velsol-paras,ABC}
 %\tag{\ref{Eq:2dmodel_ctl_Horizontal}}
 \end{subeqnarray}
And the solution is
\begin{subeqnarray}
V_r&=(A_r\cos\phi-A_i\sin\phi)re^{\psi},\\
V_{\theta}&=(B_r\cos\phi-B_i\sin\phi)re^{\psi},\\
V_z&=-2(C_r\cos\phi-C_i\sin\phi)e^{\psi},
 \label{Eq:Axisflow-velsol-vel,gamma!=0}
 %\tag{\ref{Eq:2dmodel_ctl_Horizontal}}
 \end{subeqnarray}
where $\psi=k_rz-\frac{1}{8}\gamma_rr^2$ and
$\phi=k_iz-\frac{1}{8}\gamma_ir^2$. Although either $k$ or
$\gamma$ might be complex number, only some especial combinations
could be the solutions. The first two types require $\gamma_i=0$,
in one case the magnitude of the circular velocity $V_{\theta}$
can be either less or larger that of the radial velocity $V_r$,
and in another case the magnitude of $V_{\theta}$ is larger than
that of $V_r$. Both suit for typhoon structure. The last two types
require $\gamma_r=0$, where the magnitude of $V_{\theta}$ is less
than that of $V_r$.

\subsubsection{$k_r\neq0$, $k_i=0$ and $\gamma_i=0$ }

\begin{figure}
  \centerline{\includegraphics[width=3cm,height=3cm]{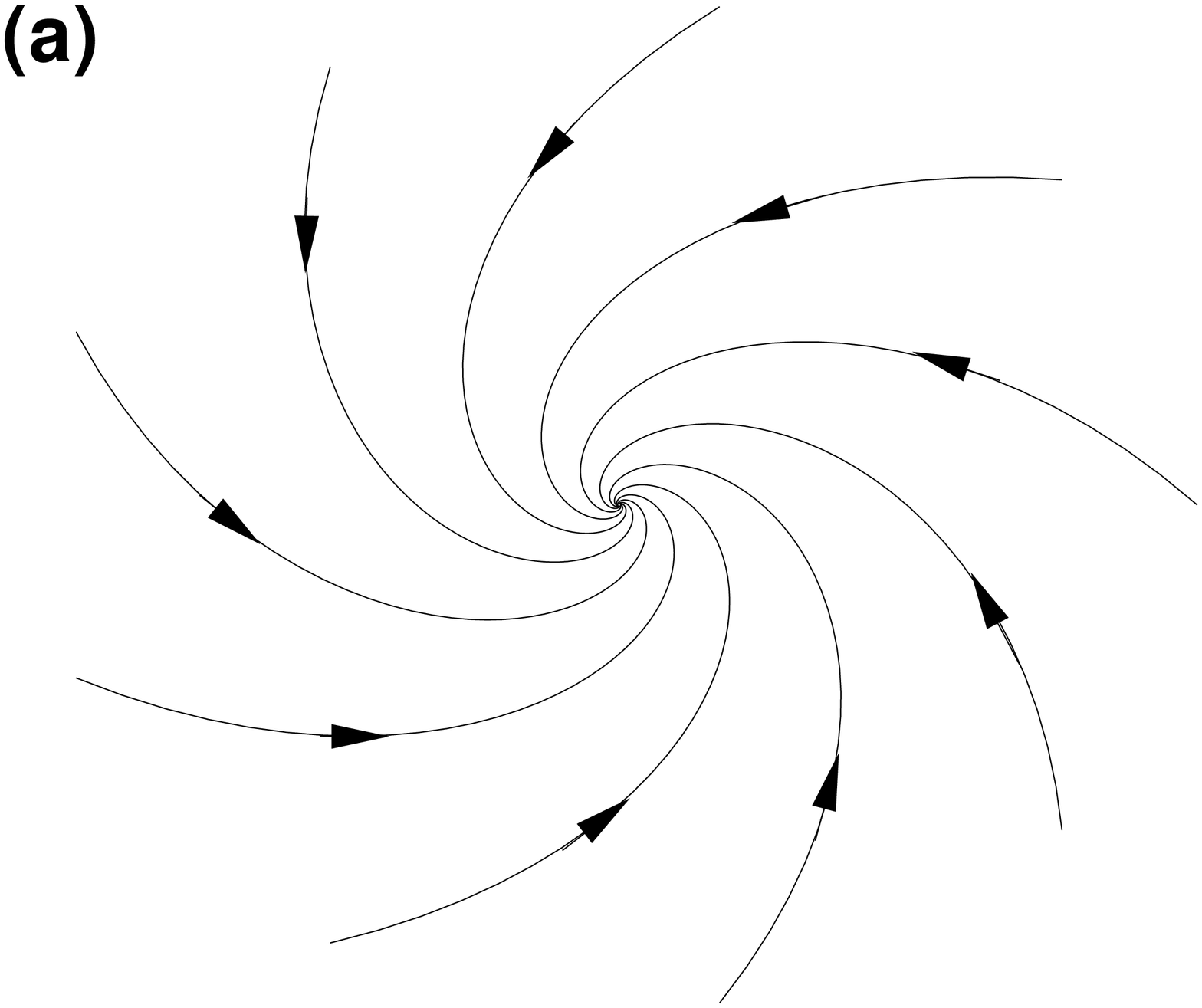}
  \includegraphics[width=3cm,height=3cm]{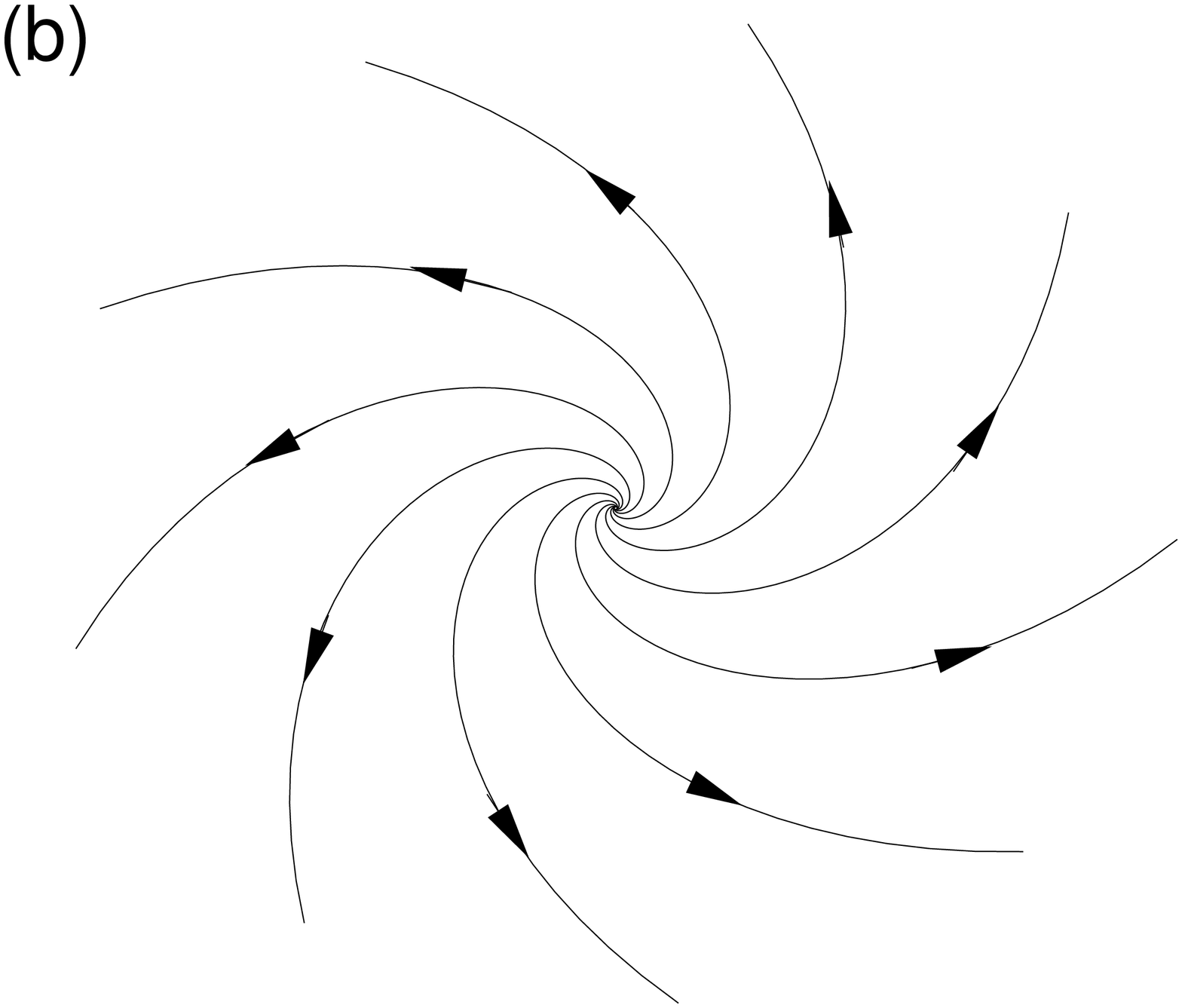}
  \includegraphics[width=3cm,height=3cm]{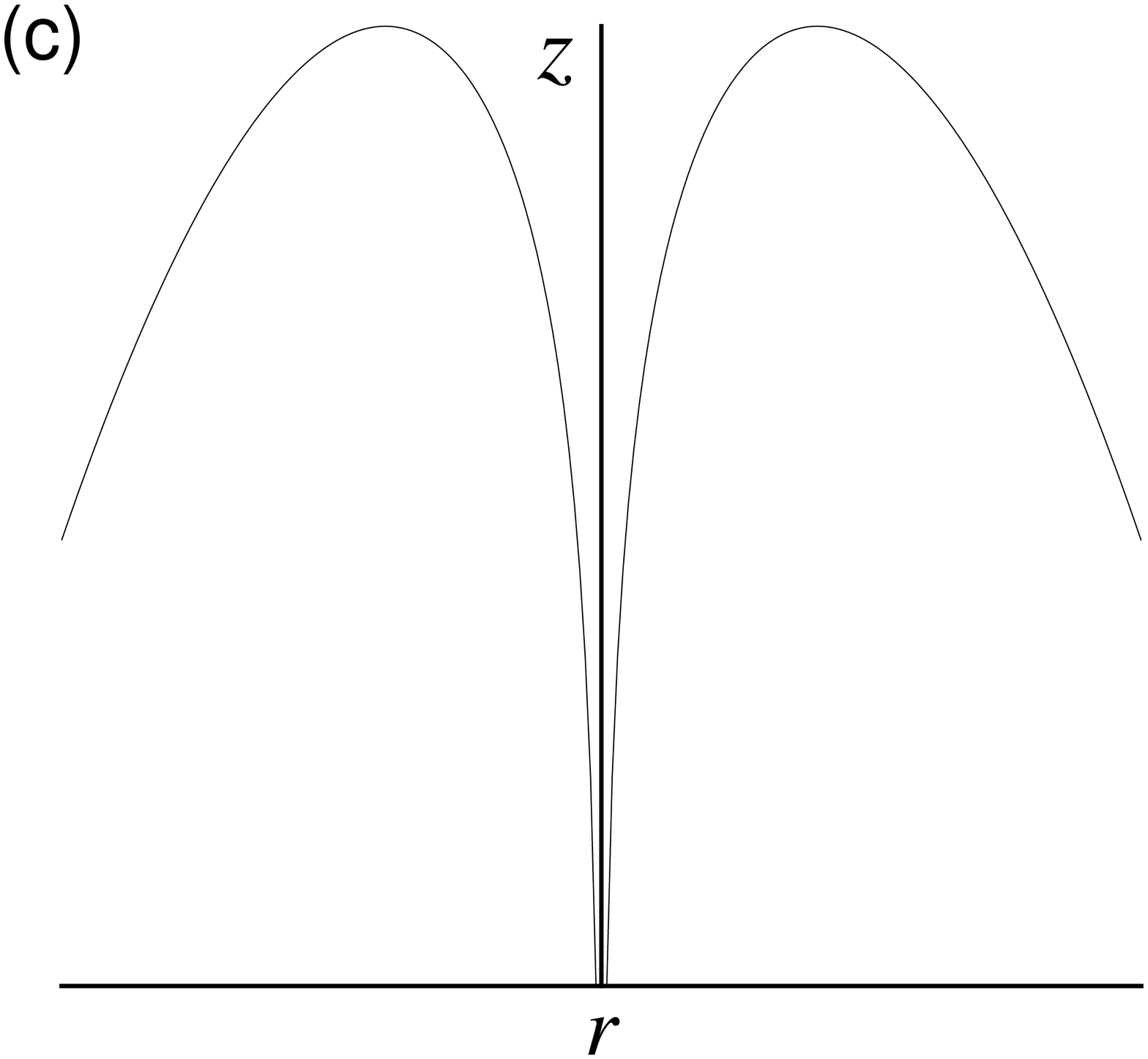}
   \includegraphics[width=3cm,height=3cm]{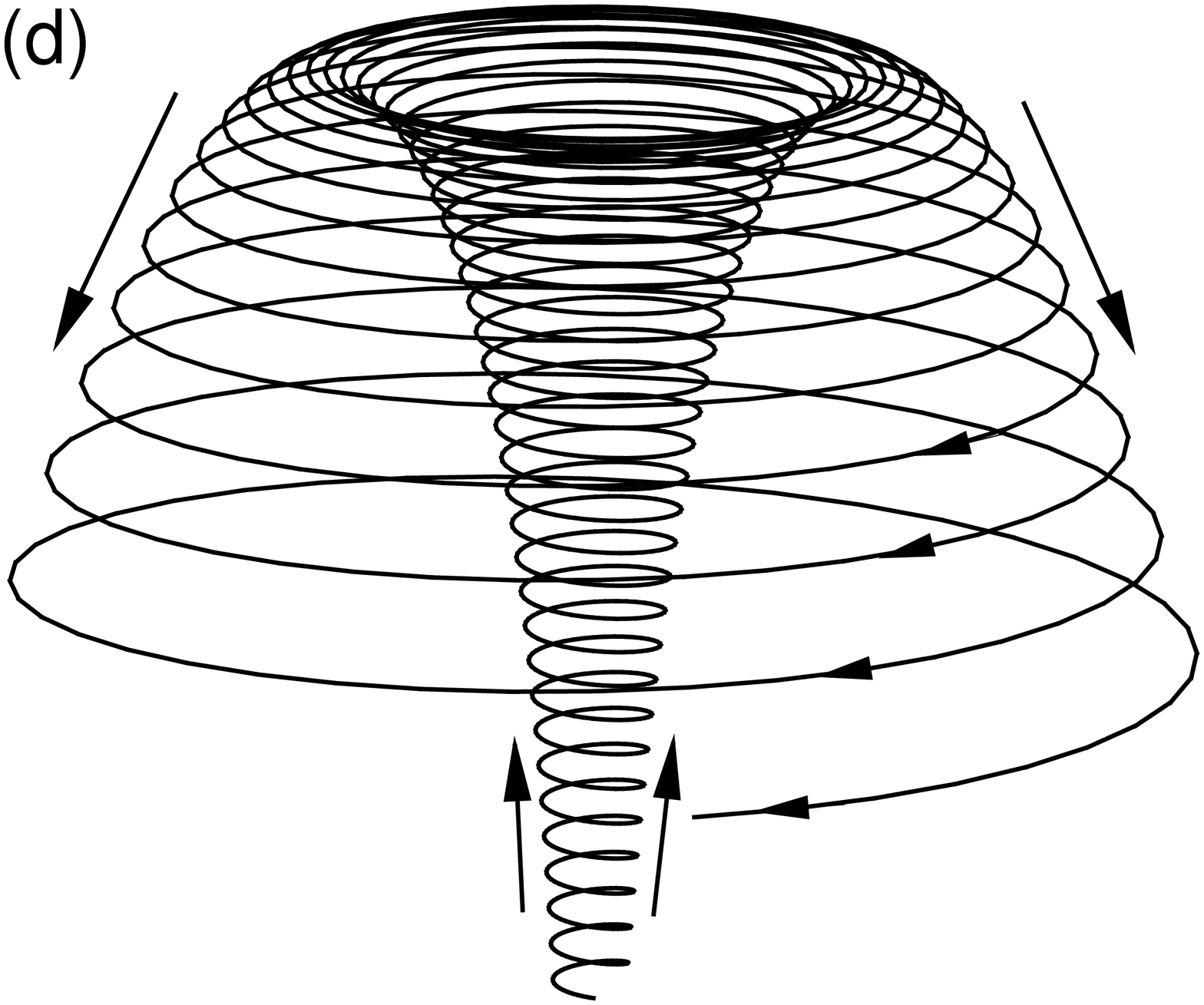}}
\caption{Paths of the fluid material elements for the umbrella
vortex.} \label{Fig:Spiral-solution-graph,gamma_i=0}
\end{figure}

In this case $\gamma$ is a real number with $\gamma_r>0$, and
$k_i=0$ implies a exponential changes in $z$,
\begin{subeqnarray}
\displaystyle V_r&=& k_rre^{k_rz-\frac{1}{8}\gamma_rr^2},\\
\displaystyle V_{\theta}&=&\lambda_r re^{k_rz-\frac{1}{8}\gamma_rr^2},\\
\displaystyle
V_z&=&-2(1-\frac{1}{8}\gamma_rr^2)e^{k_rz-\frac{1}{8}\gamma_rr^2}.
 \label{Eq:Axisflow-velsol-vel,gamma!=0,ki=0}
 %\tag{\ref{Eq:2dmodel_ctl_Horizontal}}
 \end{subeqnarray}
The solution is finite within the whole domain, if we choose
appropriate $k_r>0$ for $z<0$ and $k_r<0$ for $z>0$. The circular
velocity $V_{\theta}$ is the same with that of the Taylor vortex
for any fixed $z$. The circular velocity  maximum is $V_{\theta
m}=\lambda_r r_m e^{k_rz-1/2}$ at $r_m^2=4/\gamma_r$. And the
vertical velocity $V_z$ changes it direction at
$r_k^2=8/\gamma_r$, which we call as the radius of vortex kernel
$r_k$. Such vertical velocity distribution is something like that
of the Sullivan vortex \cite[]{TongBGVortexBook2009,WuJZbook2006}.
The paths of the fluid material elements are illuminated in Fig
\ref{Fig:Spiral-solution-graph,gamma_i=0}. In $r-\theta$ plane,
the paths are cyclonic logarithmic spirals
(Fig.\ref{Fig:Spiral-solution-graph,gamma_i=0}a) or anticyclonic
logarithmic spirals
(Fig.\ref{Fig:Spiral-solution-graph,gamma_i=0}b) with $\ln r=k_r
\theta/\lambda_r $ . In $z-r$ plan
(Fig.\ref{Fig:Spiral-solution-graph,gamma_i=0}c), the fluid
material element moves spirally along the surface decided by
$z=(\frac{1}{8}\gamma_rr^2-2\ln r)/k_r+const.$ Figure
\ref{Fig:Spiral-solution-graph,gamma_i=0}d also shows the path in
the 3D space. In the inner region $r<r_k$, the flow cyclonically
ascends from below to above as shown by upward arrow. In the outer
region $r>r_k$, the descends apart from the kernel. So the mass
conserves in this case.
\begin{equation}
\frac{2\pi r_k^2}{e} e^{k_rz}=-\int_{0}^{r_k} 2\pi r V_z dr
=\int_{r_k}^{\infty} 2\pi r V_z dr
 \label{Eq:Axisflow-velsol-gen-ekz,massmatch}
 %\tag{\ref{Eq:2dmodel_ctl_Horizontal}}
 \end{equation}

There are two intrinsic length scale in the solution. One is the
radius of maximum circular velocity $r_m$, the other is the radius
of the vortex kernel $r_k=\sqrt{2}r_m$. According to the shape of
the paths, we called it the umbrella vortex. This new vortex
solution is very like that of the typhoon structure, which we know
quite limited about.

\subsubsection{$k_r=0$, $k_i\neq0$ and $\gamma_i=0$}

In this case $\gamma$ is a real number with $\gamma_r>0$
($\lambda_r^2>k_i^2$), and $k_r=0$ implies a sinusoid function for
$z$ by taking $k_i>0$ without loss of generality,
\begin{subeqnarray}
V_r&=&-k_i\sin(k_iz)re^{-\frac{1}{8}\gamma_rr^2},\\
V_{\theta}&=&\lambda_r\cos(k_iz) re^{-\frac{1}{8}\gamma_rr^2},\\
V_z&=&-2(1-\frac{1}{8}\gamma_rr^2)\cos(k_iz)e^{-\frac{1}{8}\gamma_rr^2}.
 \label{Eq:Axisflow-velsol-vel,gamma!=0,kr=0}
 %\tag{\ref{Eq:2dmodel_ctl_Horizontal}}
 \end{subeqnarray}
Similar to the above solution,
Eq.(\ref{Eq:Axisflow-velsol-vel,gamma!=0,kr=0}) is finite within
the whole domain. But the present solution has multiply layers by
noting that the solution is periodic in $z$-coordinate. The fluid
material elements are restricted within different vertical layers.
So we call such flow as multi-planar flow. In each layer (e.g.,
$-\frac{\pi}{2}\leq k_iz\leq\frac{\pi}{2}$), the flow has a
similar behavior like that in
Eq.(\ref{Eq:Axisflow-velsol-vel,gamma!=0,ki=0}), except for that
there are both inflow ($k_iz>0$) and outflow ($k_iz<0$) at present
solution. A similar flow path of the fluid element can be find in
Fig.\ref{Fig:Spiral-solution-graph,gamma_i=0}. There are also two
intrinsic length scale in the solution. One is the radius of
maximum circular velocity $r_m$, the other is the radius of the
vortex kernel $r_k=\sqrt{2}r_m$. Another notable property of the
multi-planar solution is that there might be numberless solutions
for a given boundary condition at $z$-coordinate. So the solution
of the 3D Euler equations for a given boundary condition might not
be uniqueness.

\subsubsection{$k_r=0$, $k_i=\pm \lambda_r$ and $\gamma_r=0$ }
\begin{figure}
  \centerline{\includegraphics[width=3cm]{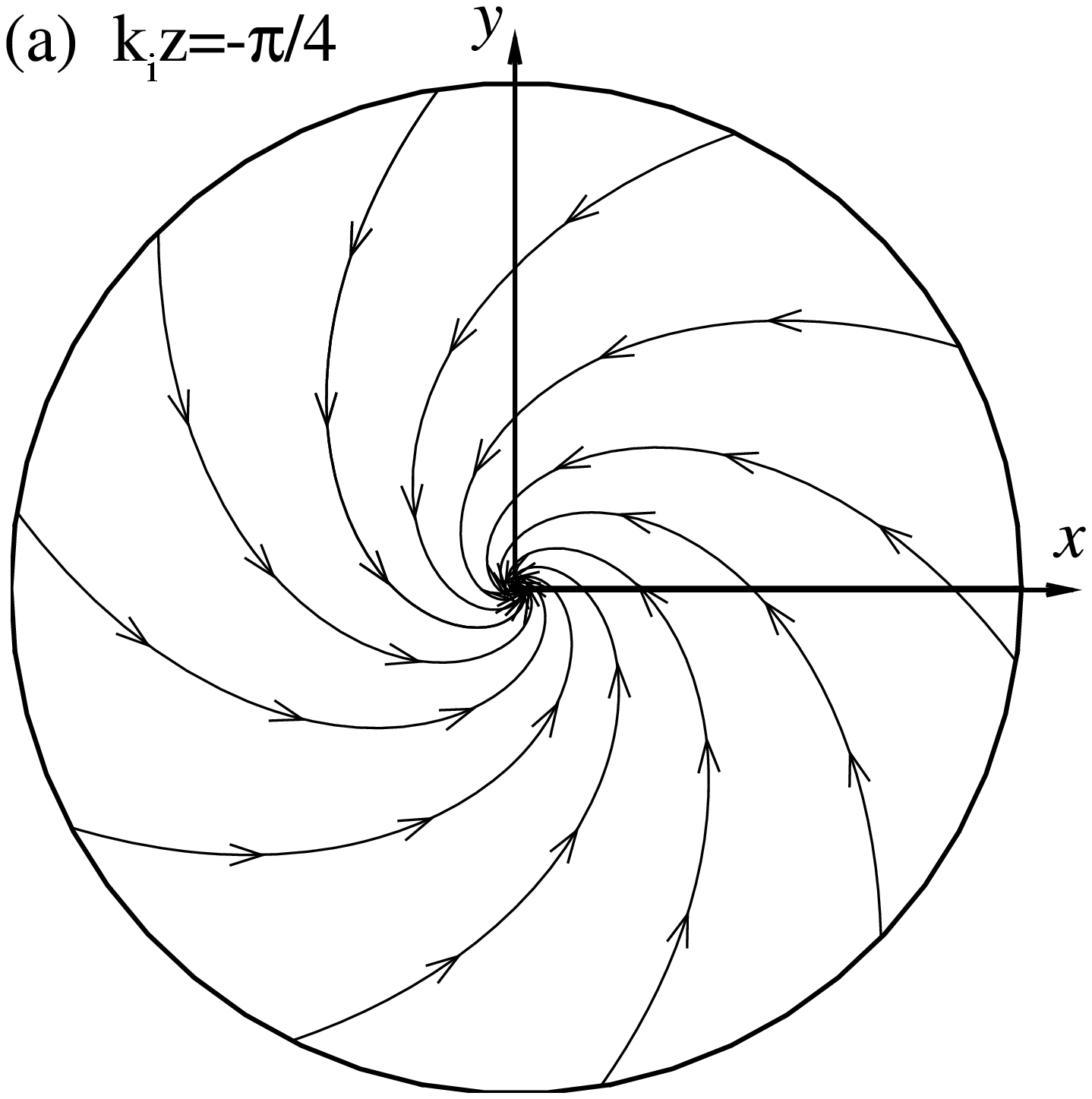}
  \includegraphics[width=3cm]{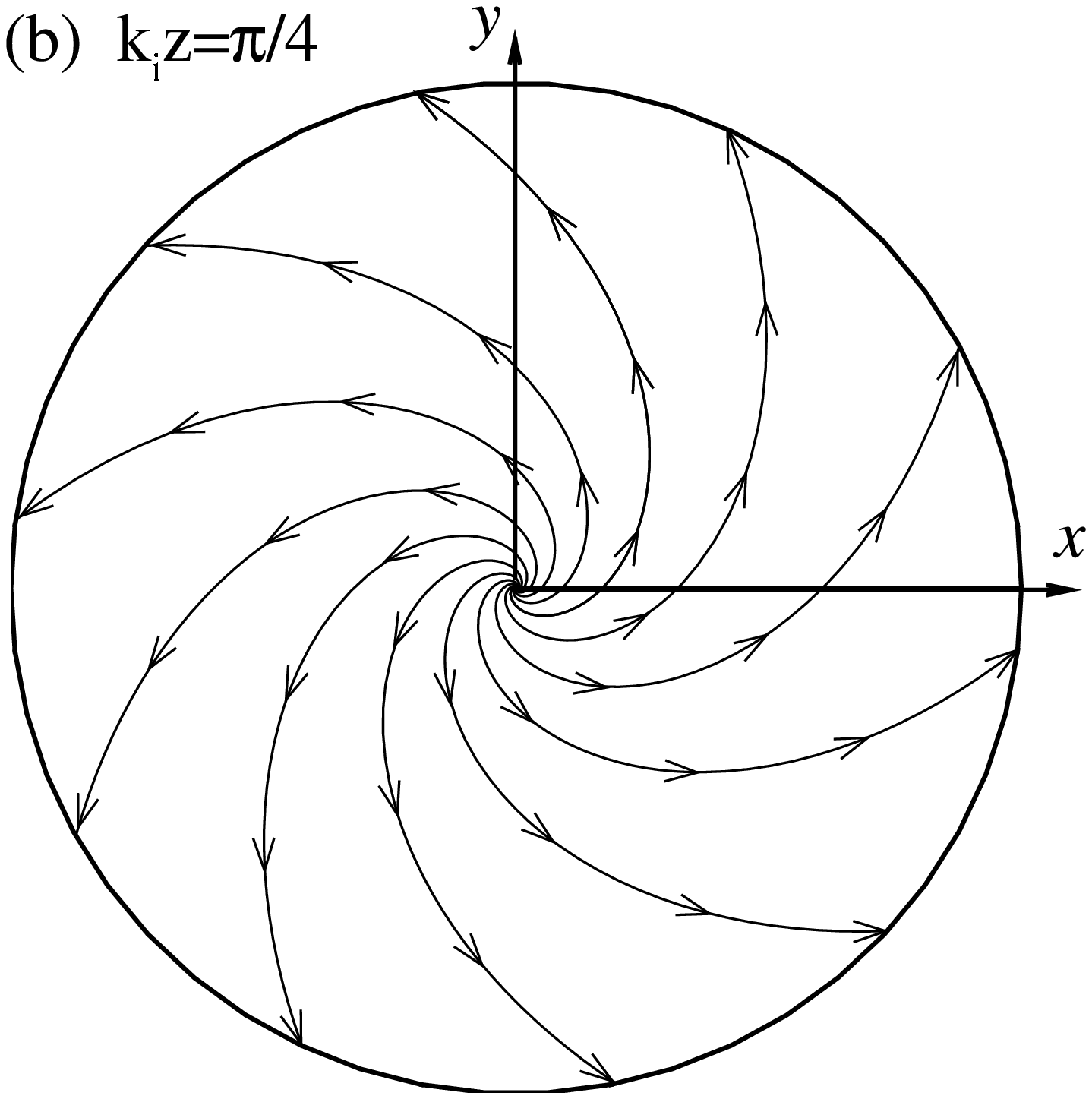}
  \includegraphics[width=3cm]{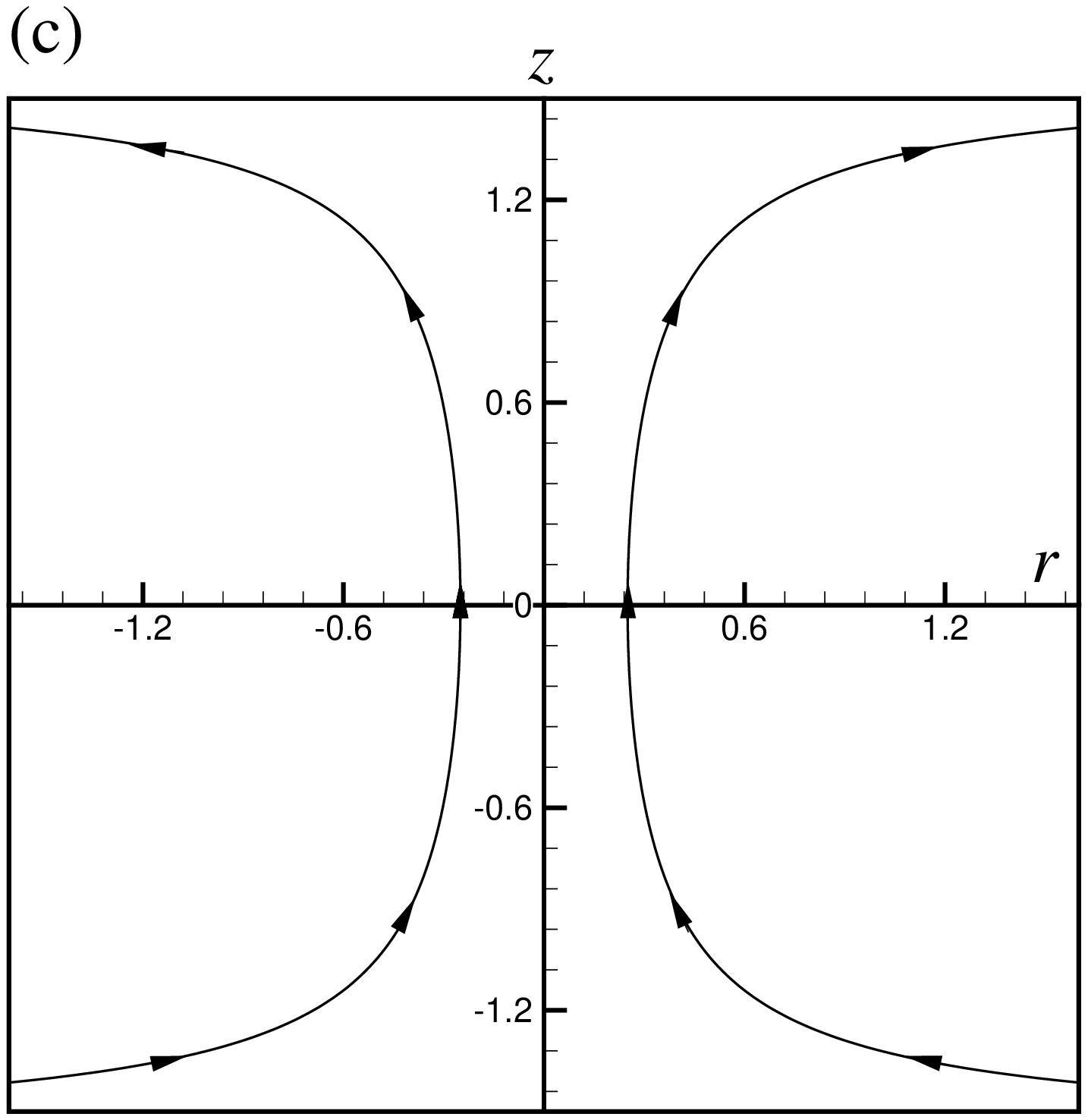}
   \includegraphics[width=3cm]{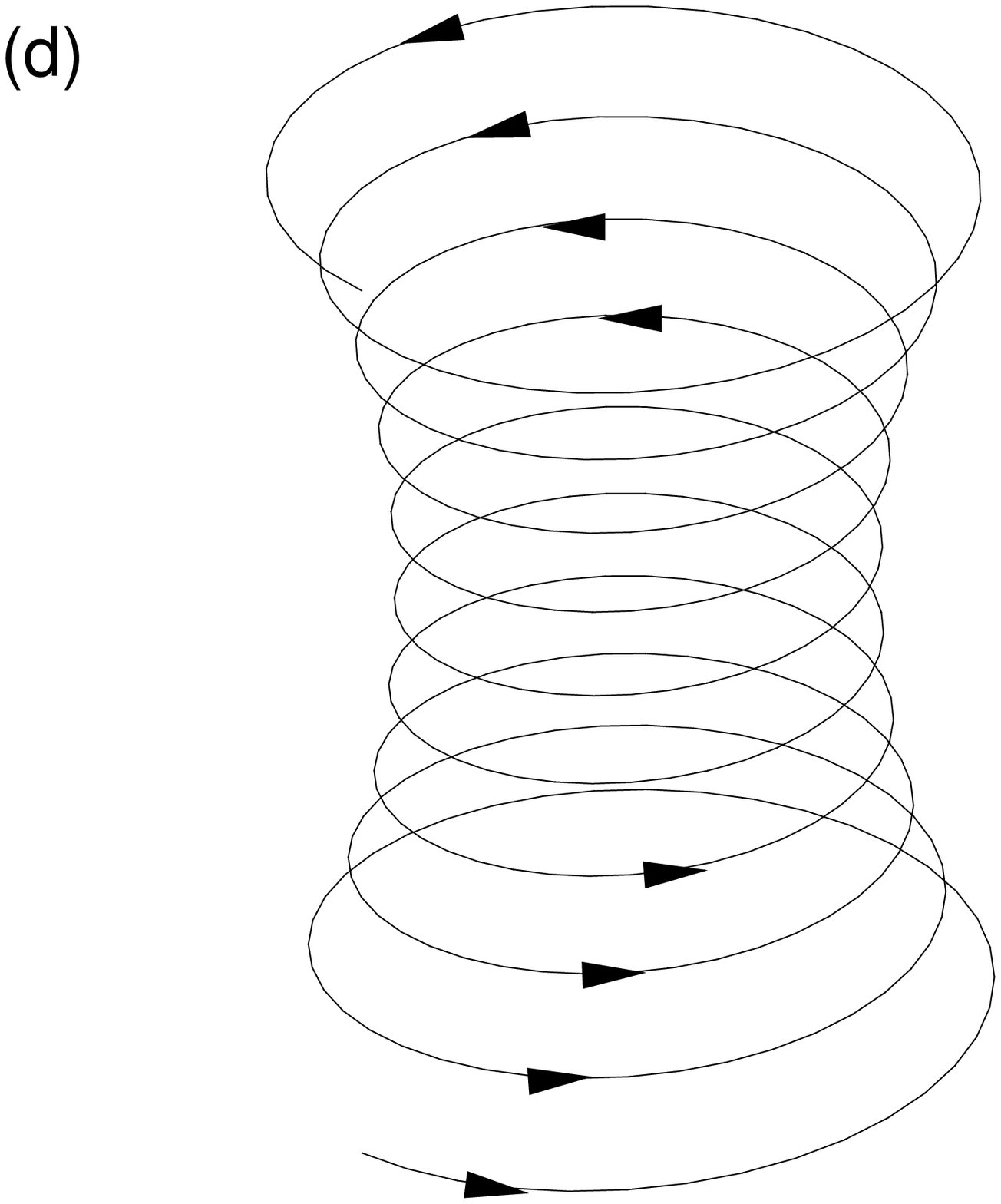}}
\caption{Paths of the fluid material elements for multi-planer
vortex within a layer.} \label{Fig:Spiral-solution-graph,gamma=0}
\end{figure}

In this case, $\gamma_r=\gamma_i=0$, the magnitude of $V_r$ is the
same with that of $V_{\theta}$ (like Ekman spiral). And the
solution of $R$ is Eq.(\ref{Eq:Axisflow-velsol-R,gamma=0}) for
$\gamma=0$. In Eq.(\ref{Eq:Axisflow-velsol-R,gamma=0}a), the
solution is finite for inner kernel $r<r_k$ but infinite for far
region $r\rightarrow\infty$.  While in
Eq.(\ref{Eq:Axisflow-velsol-R,gamma=0}b), the solution is finite
for far region $r>r_k$ but infinite for $r\rightarrow0$. Similar
to the solution of the Rankine vortex, we can combine the above
solutions to obtain a uniform solution at the interface $r=r_i$,
which is finite at whole region. However, either radial velocity
$V_r$ or the vertical velocity $V_z$ might be discontinued in
Eq.(\ref{Eq:Axisflow-velsol-gen-gamma=0,kr=0}). Additional
notation for $r_k$ will be given in
~\S\,\ref{sec:sepcialsol-keq0-H=z}.

For example, we choose $V_r$ to be continued at the interface
$r_i^2=e^{-cr_i^2}=1/c$, hence the solution is,
\begin{subeqnarray}
&V_r = k_ir\sin(k_iz), \,\,V_{\theta} =\pm k_i r \cos(k_iz), \,\, V_{z}=2\cos(k_iz)\\
&V_r =\displaystyle \frac{k_i}{r}e^{-cr^2}\sin(k_iz),
\,\,V_{\theta} =\pm \frac{k_i}{r}e^{-cr^2}\cos(k_iz), \,\, V_{z} =
-2ce^{-cr^2}\cos(k_iz).
 \label{Eq:Axisflow-velsol-gen-gamma=0,kr=0}
 %\tag{\ref{Eq:2dmodel_ctl_Horizontal}}
 \end{subeqnarray}
The present solution has same function as
Eq.(\ref{Eq:Axisflow-velsol-vel,gamma!=0,kr=0}) in $z$-coordinate.
Besides, the velocities change their directions along the
$r$-coordinate, which is like that in
Eq.(\ref{Eq:Axisflow-velsol-vel,gamma!=0,kr=0}). Figure
\ref{Fig:Spiral-solution-graph,gamma=0} shows the path of a fluid
material element within $-\pi/2\leq k_iz\leq\pi/2$ for
Eq.(\ref{Eq:Axisflow-velsol-gen-gamma=0,kr=0}a). According to
Eq.(\ref{Eq:Axisflow-velsol-path}), in $r-\theta$ plan, the paths
are logarithmic spirals $\ln r=\tan (k_iz) \theta$
(Fig.\ref{Fig:Spiral-solution-graph,gamma=0}a,b). In $z-r$ plan
(Fig.\ref{Fig:Spiral-solution-graph,gamma=0}c), the fluid material
element moves spirally along the surface decided by
$\cos(k_iz)r^2=const.$ Figure
\ref{Fig:Spiral-solution-graph,gamma=0}d also shows the path in
the 3D space. It should be noted that the flow field is
respectively inflow and outflow at lower part $-\pi/2\leq k_iz
\leq0$ (Fig.\ref{Fig:Spiral-solution-graph,gamma=0}a) and higher
part $0\leq k_iz \leq \pi/2$
(Fig.\ref{Fig:Spiral-solution-graph,gamma=0}b). However, the path
 of the fluid material element is always cyclonic from below to
 above (Fig.\ref{Fig:Spiral-solution-graph,gamma=0}d).

The radial velocity $V_r$ and the circular velocity $V_{\theta}$
continue, but the vertical velocity $V_z$ discontinue at $r=r_i$.
If the flow within the inner kernel ascends at $r<r_i$, then the
flow descends at $r>r_i$. The total updraft mass is equal to the
total downdraft mass.
\begin{equation}
2\pi r_i^2 \cos(k_iz)=\int_{0}^{r_i} 2\pi r V_z dr
=-\int_{r_i}^{\infty} 2\pi r V_z dr
 \label{Eq:Axisflow-velsol-gen-ecos,massmatch}
 %\tag{\ref{Eq:2dmodel_ctl_Horizontal}}
 \end{equation}

\subsubsection{$k_r\neq0$, $k_i\neq0$ and $\gamma_r=0$}
\begin{figure}
  \centerline{\includegraphics[width=3cm,height=3cm]{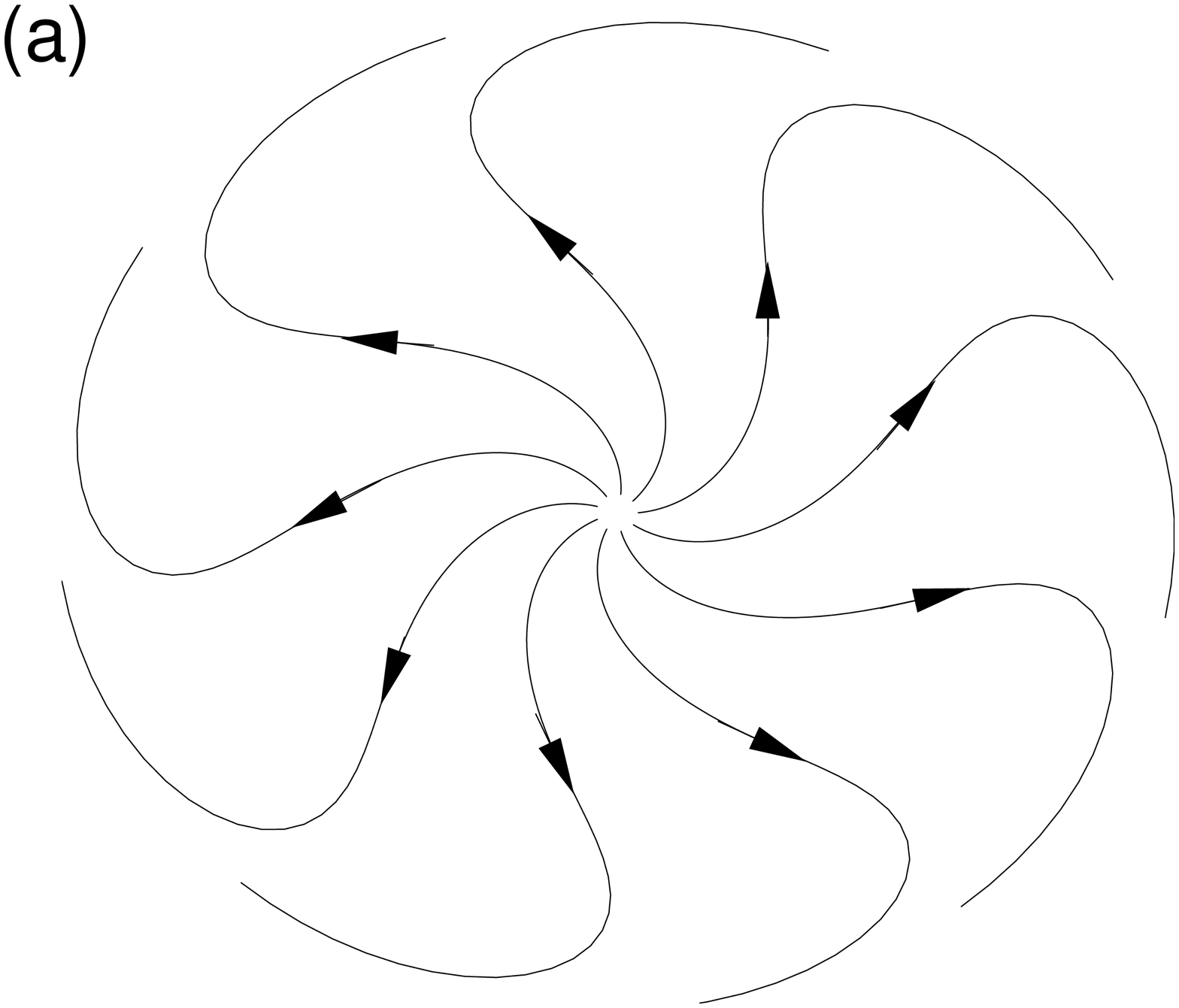}
  \includegraphics[width=3cm,height=3cm]{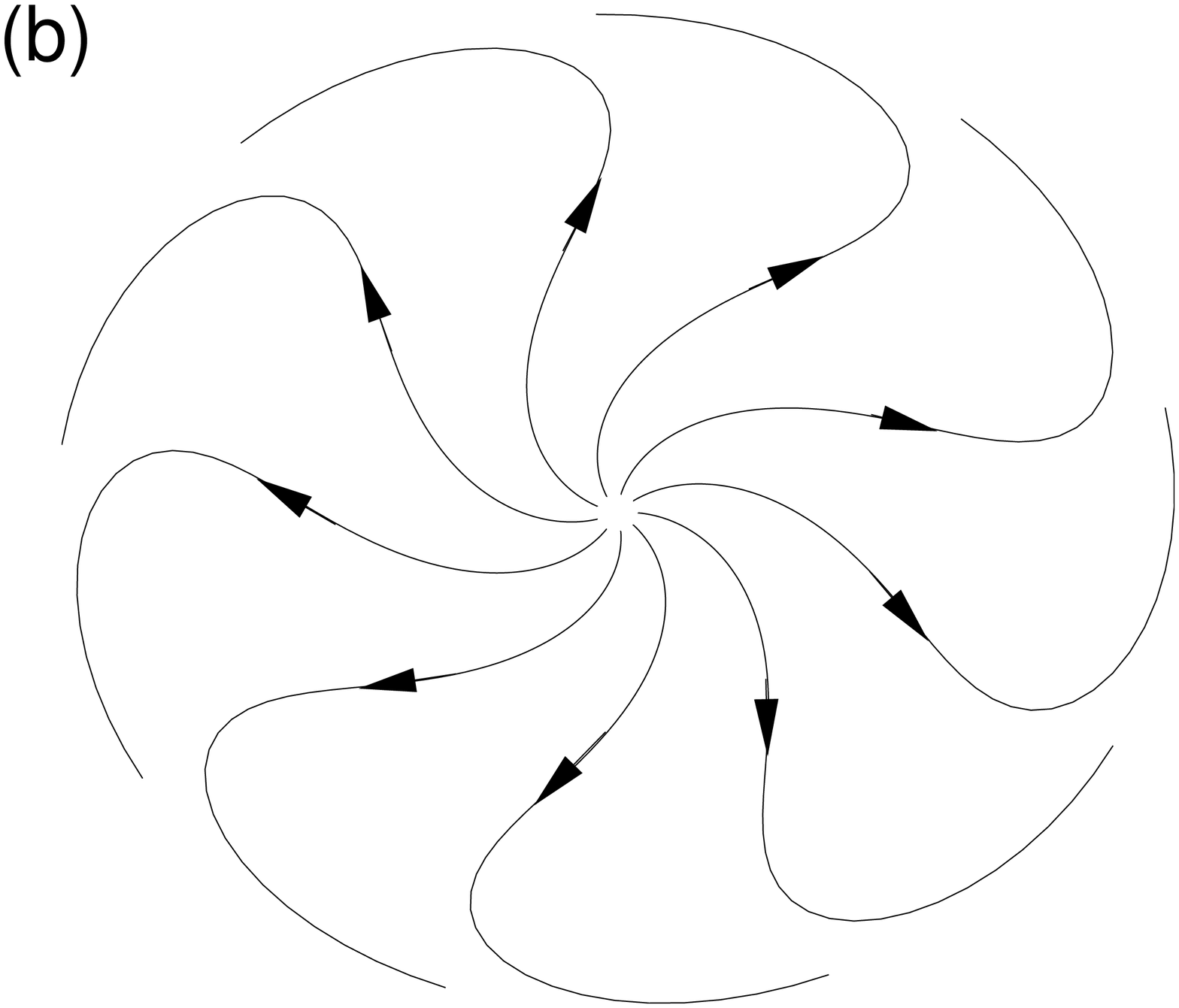}
  \includegraphics[width=3cm,height=3cm]{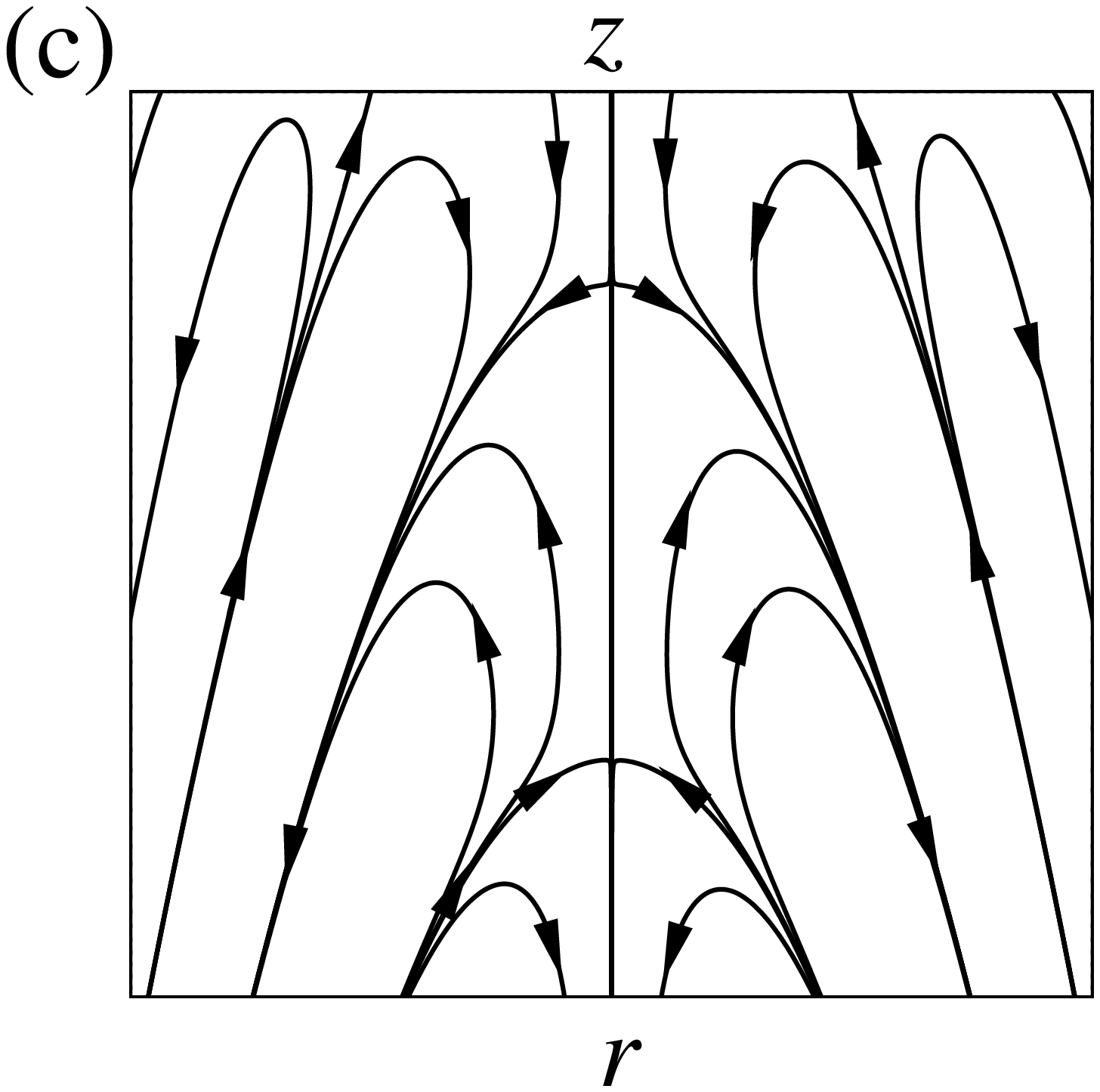}
   \includegraphics[width=3cm,height=3cm]{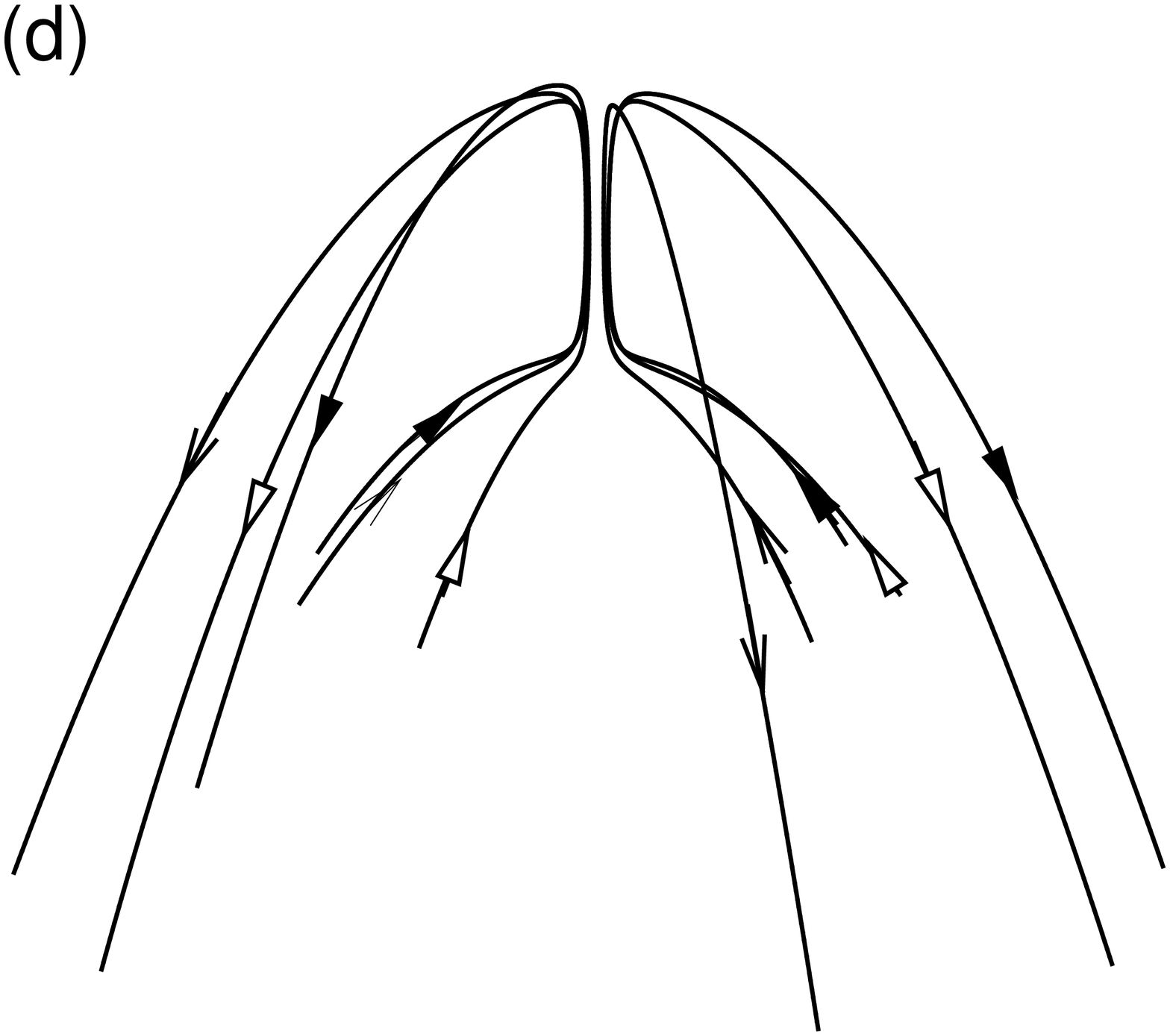}}
\caption{Paths of the fluid material elements for multi-paraboloid
vortex.} \label{Fig:Spiral-solution-graph,gamma_r=0}
\end{figure}

In this case $\gamma$ is a pure image number, and $\gamma_r=0$
implies $k_r^2-k_i^2+\lambda_r^2=0$. The solution is,
\begin{subeqnarray}
V_r&=&[k_r\cos(k_iz-\frac{1}{8}\gamma_ir^2)-k_i\sin(k_iz-\frac{1}{8}\gamma_ir^2)]re^{k_rz},\\
V_{\theta}&=&\lambda_r\cos(k_iz-\frac{1}{8}\gamma_ir^2) re^{k_rz},\\
V_z&=&-2[\cos(k_iz-\frac{1}{8}\gamma_ir^2)+\frac{\gamma_i}{8}r^2\sin(k_iz-\frac{1}{8}\gamma_ir^2)]e^{k_rz}.
 \label{Eq:Axisflow-velsol-vel,gamma!=0,gr=0}
 %\tag{\ref{Eq:2dmodel_ctl_Horizontal}}
 \end{subeqnarray}
It is obvious that this solution can only be an inner solution
$r<r_k$, as they might be infinity for $r\rightarrow\infty$. In
this case, the magnitude of $V_r$ is $\sqrt{k_r^2+k_i^2}$, which
is larger than that of $V_{\theta}$. So the rotation is relatively
small, which can be seen from
Fig.\ref{Fig:Spiral-solution-graph,gamma_r=0}a,b. Both the radio
velocity and azimuth velocity may change directions along the
$r$-coordination and $z$-coordination. The solution is also a
multi-paraboloid vortex by noting that the velocities are
spatially periodic shift along the paraboloid of
$k_iz=\frac{1}{8}\gamma_ir^2$
(Fig.\ref{Fig:Spiral-solution-graph,gamma_r=0}c). For example,
Fig.\ref{Fig:Spiral-solution-graph,gamma_r=0}d depicts the the
paths of the fluid material elements within a paraboloid layer.
Due to the smaller azimuth velocity, the paths only have some
slight twists, whereas there are lots of spirals in
Fig.\ref{Fig:Spiral-solution-graph,gamma_i=0}.

\subsection{$k=0$}

In this case, let $H(z)=z^n, n=0,1,2$. As the solution may be
infinity for $n=1,2$ as $z\rightarrow \infty$, so it should be
taken only as a solution in lower layer $z<\infty$.

\subsubsection{$H(z)=1$}\label{sec:sepcialsol-keq0-H=1}
Firstly, considering $n=0$ and $H(z)=1$, the radial velocity $V_r$
vanishes according to Eq.(\ref{Eq:Axisflow-velsol-gen}). This is
barotropic flows in the geophysics, as the solution is free of
$z$-coordinate. Any differentiable function $R(r)$, the circular
velocity $V_{\theta}$ and vertical velocity $V_z$ would be the
solution from Eq.(\ref{Eq:Axisflow-velsol-cross}) due to that
$V_{\theta}$ and $V_z$ are decoupled for $H(z)=1$, which is known
as a stretch-free inviscid vortex can have arbitrary radial
dependence \cite[]{WuJZbook2006}. In present solution,
$V_{\theta}$ and $V_z$ does not fully decoupled, as
Eq.(\ref{Eq:Axisflow-velsol-genH=1}) shows.
\begin{equation}
V_r=0, \, \, V_{\theta} = \lambda \frac{R(r)}{r}, \,\,V_{z} = -
\frac{R'(r)}{r}
 \label{Eq:Axisflow-velsol-genH=1}
 %\tag{\ref{Eq:2dmodel_ctl_Horizontal}}
 \end{equation}
Given $R(r)=1$ is the line vortex, $R(r)=r^2$ is the solid
rotation. And the Rankine vortex and the Taylor vortex can be
obtained by given $R(r)=ar^2+b$ and $R(r)=r^2e^{-ar^2}$,
respectively. The Batchelor vortex and Oseen-Lamb vortex can also
be obtained by given $R(r)=1-e^{-r^2}$, etc \cite[]{WuJZbook2006}.

Although any differentiable function $R(r)$ gives the vortex
solution, not all the vortex can be hold due to the instabilities,
including the centrifugal instability and the shear instability
\cite[]{CriminaleBook2003,WuJZbook2006,SunL2006c}. In
\cite{SunL2006c}, a general stability criterion was obtained for
such circular flows. It was found that the longwaves (e.g.
wavenumbers are 1, 2, etc.) might be unstable and break into
asymmetrical vortices if there is a local maximum in vorticity
distribution along the radial coordinate. The numerical
simulations also approved this
\cite[]{Roger1990,Belotserkovskii2009}.

\subsubsection{$H(z)=z$} \label{sec:sepcialsol-keq0-H=z}

Secondly, consider $n=1$ and $H(z)=z$. If $\lambda \neq 0$, the
solution is,
\begin{equation}
V_r=a re^{-\frac{1}{8}\lambda^2r^2},\,\, V_{\theta}=a\lambda
re^{-\frac{1}{8}\lambda^2r^2}z,
\,\,V_z=-2a(1-\frac{1}{8}\lambda^2r^2)e^{-\frac{1}{8}\lambda^2r^2}z
\label{Eq:Axisflow-velsol-genH=z,lambdaneq0}
 \end{equation}
And if $\lambda=0$, the circular velocity $V_{\theta}$ vanish due
to this. Similar to
Eq.(\ref{Eq:Axisflow-velsol-gen-gamma=0,kr=0}), the present
solution Eq.(\ref{Eq:Axisflow-velsol-genH=z}) is also the
combination of two parts at $r=r_i$, where a discontinuity might
occur. One should note that we have chosen $a=b=1$ and $c=1/r_i^2$
in Eq.(\ref{Eq:Axisflow-velsol-gen-gamma=0,kr=0}). Hence, the
solution for for the whole domain is,
\begin{subeqnarray}
V_r=ar,\,\, &V_{\theta}=0, \,\,&V_z=-2az \\
V_r = \frac{b}{r}e^{-cr^2}, \,\,&V_{\theta} = 0, \,\, &V_{z} =
2bcze^{-cr^2}\label{Eq:Axisflow-velsol-genH=z}
 \end{subeqnarray}
The vertical velocity $V_z$ discontinue at $r=r_i$, and the
vertical velocity discontinuity is $\triangle V_z=2a(1+br_i^2)$.
Similarly, we can choose $a=-bce^{-cr_i^2}$, thus the vertical
velocity is continued but the radial velocity has a discontinuity
of $\triangle V_r=a(c/r_i+r_i)$, which is also a two-cell vortex.

\subsubsection{$H(z)=z^2$}
Finally, consider $n=2$ and $H(z)=z^2$. As $\gamma=0$ implies
$\lambda=0$, the circular velocity $V_{\theta}$ vanish due to
this. Only $R=r^2$ could satisfy
Eq.(\ref{Eq:Axisflow-velsol-cross}), and the solution is,
\begin{equation}
V_r=2rz,V_{\theta}=0, V_z=-2z^2
\label{Eq:Axisflow-velsol-genn=2-Insol}
 \end{equation}
This is the solution of rotational stagnation flow over a plate
\cite[]{WangCY1991}, or the Hill spherical vortex
\cite[]{Batchelor1967,TongBGVortexBook2009,WuJZbook2006}.

\section{Discussion}\label{sec:discussion}

\subsection{Extensions of Solutions}
In above section, we assumed $b=0$ in
Eq.(\ref{Eq:Axisflow-velsol-R,gamma=0}a) to obtain the inner
solution. Alternatively, we can take $a=0$ and $b\neq0$ to as a
outer solution, which is a well-known pure vortex solution for
stretch-free vortex \cite[]{WuJZbook2006}. It is obvious that
Eq.(\ref{Eq:Axisflow-velsol-R,gamma=0}a) is the solution of the
Couette-Taylor flow, we can also use
Eq.(\ref{Eq:Axisflow-velsol-R,gamma=0}a) and $H(z)$ to obtain new
solutions.

In Eq.(\ref{Eq:Axisflow-velsol-vel,gamma!=0,kr=0}) and
Eq.(\ref{Eq:Axisflow-velsol-gen-gamma=0,kr=0}), both solutions
have a same $z$-coordinate dependence function, so these solutions
can also be combined for some new solutions. For example, if we
take Eq.(\ref{Eq:Axisflow-velsol-vel,gamma!=0,kr=0}) for $r<r_i$
and Eq.(\ref{Eq:Axisflow-velsol-gen-gamma=0,kr=0}b) for $r>r_i$ as
a new solution, where $r_i^2=1/(\frac{\gamma_r}{8}-1)$ and
$c=(1-\frac{\gamma_r}{8})/(3+\frac{\gamma_r}{8})$ in
Eq.(\ref{Eq:Axisflow-velsol-gen-gamma=0,kr=0}b), both the $V_r$
and $\partial V_r/\partial r$ of the new combined solution are
continued at $r=r_i$. As this new solution is continued, it might
be better than that of discontinued solution in
Eq.(\ref{Eq:Axisflow-velsol-gen-gamma=0,kr=0}). It is noted that
the value of $\gamma_r$ could be negative in the near-kernel
region for present solution, although the solution of
Eq.(\ref{Eq:Axisflow-velsol-vel,gamma!=0,kr=0}) requires
$\gamma_r>0$ in the far region. The negative $\gamma_r$ implies a
very fast tangential velocity rise in the near-kernel region,
which is true in the typhoon.

According to the previous studies
\cite[]{Batchelor1967,Frewer2007}, the solution set is quite
large, which can also be seen from the
Eq.(\ref{Eq:Axisflow-velsol-gen}). According to
\cite{Batchelor1967}, the vertical velocity
$V_{\theta}=C(\Psi)/r$, while we simply took the velocity
components as a form  $V_{\theta}=\lambda\Psi/r$ in
Eq.(\ref{Eq:Axisflow-velsol-gen}). Beyond present study, there
should be other axisymmetric solutions. For example, we can apply
the same approach to find other exact solutions by taking
$V_{\theta}=\lambda/r$ or $V_{\theta}=\lambda \Psi^2(r,z)/r$, etc.
The method used in present work could also be applied to other
complex flows, e.g., the geophysical flow in a rotating frame
($f$-plane), the magnetohydrodynamics (MHD) in astrophysics, and
even for the viscous flows. According to \cite{WangCY1991}, the
generalized Beltrami flow might also be the steady solution of
Navier-Stokes equations, if the curl of vorticity is potential.
Moreover, the present solutions might further be used to obtain
non-steady solutions of Navier-Stokes equations, like that of
Oseen-Lamb vortex.

\subsection{Possible Applications}

As mentioned above, a stretch-free inviscid vortex
(two-dimensional axisymmetric columnar vortex) can have arbitrary
radial dependence for $H(z)=1$ \cite[]{WuJZbook2006}. However,
many well-known vortex solutions are always similar to either
Eq.(\ref{Eq:Axisflow-velsol-R,gamma=0}a) or
Eq.(\ref{Eq:Axisflow-velsol-R,gamma=0}b). It is from this study
that these two-dimensional axisymmetric columnar vortices are also
the three-dimensional axisymmetric columnar vortex solutions. So
we can find either the Rankine or Taylor vortex for any fixed
layer $z=const$.

As some new exact solutions are finite within the whole region,
they can be applied to study the radial structure of the typhoons,
the tornados and the mesoscale eddies in the geophysical flows. As
mentioned above, there are 3 independent parameters in the
solution. To determinate them, we need the measurable qualities,
such as the radius of circular velocity maximum $r_m$, the maximum
circular velocity $V_{\theta m}$, the spiral angle defined by
$k/\lambda$, the descend/ascend flow mass in
Eq.(\ref{Eq:Axisflow-velsol-gen-ekz,massmatch}) and
Eq.(\ref{Eq:Axisflow-velsol-gen-ecos,massmatch}), the horizontal
inflow/outflow mass (which can also be determined by both
$k/\lambda$ and the descend/ascend flow mass), the angular
momentum $m$, etc. According to Eq.(\ref{Eq:Axisflow-velsol-gen}),
the angular momentum $m=rV_{\theta}=-\lambda R(r)H(z)$, thus the
solutions of $R(r)$ (Eq.(\ref{Eq:Axisflow-velsol-R,gamma!=0}) and
Eq.(\ref{Eq:Axisflow-velsol-R,gamma=0})) can be applied to discuss
the angular momentum of the vortex. For the typhoon observations,
such solutions can be used to fit the real velocity distribution
along the radial coordinate. This may also be useful to classify
the typhoons and the tornados according the flow structures
provided by above solutions.

\section{Conclusion}\label{sec:conclusion}

A general exact axisymmetric spiral solution was obtained for 3D
incompressible Euler equations, and some exact solutions were also
obtained. The solution describes the spiral path of the fluid
material element on the Bernoulli surface. The explicit umbrella
vortex and multi-layer vortex solutions, which are new in this
investigation, might be used to describe the 3D structure of the
tropical cyclones, tornados and mesoscale eddies in the
geophysical flows. The solutions also imply that the spiral
structure is the intrinsic structure of the flows in the nature.

The method used in present work is very straightforward, and it
could also be applied to other complex flows, e.g., the
geophysical flows in a rotating frame ($f$-plane), and even for
the non-steady viscous flows.

The author thanks Prof. Wang W. at OUC, who discussed lots of
vortex dynamic problems with the author and encouraged the author
to finish this work. The author also thanks Dr. Wang Bo-fu and Dr.
Wan Zhen-hua for their help to check the formula, Prof. Huang
Rui-Xin at WHOI for useful comments. Prof. Yin X-Y at USTC is also
acknowledged, who led the author to this field. This work is
supported by the National Basic Research Program of China (No.
2007CB816004), and the Knowledge Innovation Program of the Chinese
Academy of Sciences (Nos. KZCX2-YW-QN514).

\bibliographystyle{jfm}

%\bibliography{MSH1}

\end{document}